\documentclass[10pt,letterpaper]{article}
\usepackage{geometry}

\usepackage{changepage}
\usepackage[utf8]{inputenc}

\usepackage{textcomp,marvosym}

\usepackage{fixltx2e}

\usepackage{amsmath,amssymb}

\usepackage{cite}

\usepackage[hidelinks]{hyperref}
\usepackage{nameref}

\usepackage[right]{lineno}

\usepackage{microtype}

\usepackage{rotating}

\usepackage[aboveskip=1pt,labelfont=bf,labelsep=period,justification=raggedright,singlelinecheck=off]{caption}

\makeatletter
\renewcommand{\@biblabel}[1]{\quad#1.}
\makeatother

\date{}

\begin{document}
\vspace*{0.35in}

\begin{flushleft}
{\Large
\textbf\newline{Multifractal and Network Analysis of Phase Transition}
}
\newline
\\
Longfeng Zhao\textsuperscript{*},
Wei Li,
Chunbin Yang,
Jihui Han,
Zhu Su,
Yijiang Zou,
Xu Cai\\
\bigskip
\bf Complexity Science Center\& Institute of Particle Physics, Hua-Zhong (Central China) Normal University, Wuhan, China

\bigskip

%
%





* zlfccnu@mails.ccnu.edu.cn

\end{flushleft}
\section*{Abstract}
Many models and real complex systems possess critical thresholds at which the systems shift from 
one sate to another. The discovery of the early warnings of the systems in the vicinity of critical point
are of great importance to estimate how far a system is from a critical threshold.
Multifractal Detrended Fluctuation analysis (MF-DFA) and visibility graph method
have been employed to investigate the fluctuation and geometrical structures of magnetization time
series of two-dimensional Ising model around critical point. The Hurst exponent has been confirmed to be a good
indicator of phase transition. Increase of the multifractality of the time series
have been observed from generalized Hurst exponents and singularity spectrum. 
Both Long-term correlation and broad probability density function are identified 
to be the sources of multifractality of time series near critical regime. Heterogeneous nature of the 
networks constructed from magnetization time series have validated the fractal properties of 
magnetization time series from complex network perspective.
Evolution of the 	topology quantities such as clustering coefficient, average degree, 
average shortest path length, density, assortativity and heterogeneity serve as early warnings of phase
transition. Those methods and results can provide new insights about analysis of phase transition
problems and can be used as early warnings for various complex systems.


\section*{Introduction}
Complex systems are consisted with subunits that
interact non-linearly with each other.
Among so many general properties which describe the complex systems, the existence of
critical threshold is apparently common to plenty of complex systems\cite{Scheffer2009}.
It is not possible to fully anticipate
the behaviour of the system in terms of behaviours of its 
components. Thus characterizing the dynamical
process of a complex system from time series is a fundamental
problem of significant importance in many research fields\cite{Zhang2006,Kwapien2012}. Here
we focus on the very well known physical system, two-dimensional Ising
model\cite{Ising,Brush1967}. It is a statistical physics model for ferromagnetic materials which goes
through phase transition at non-zero temperature. The Ising model has been 
widely applied to different fields such as social systems\cite{Castellano2009} and
financial systems\cite{Sornette2014}. We use the 
Metropolis algorithm\cite{Metropolis1953}to simulate two-dimensional Ising system. The 
output of the simulation are time series of average magnetization of
the system at different temperature. Time evolution of average magnetization
can be used to reveal the dynamical properties of the system.\\

Fractals and mutifractals are ubiquitous in physics and social systems\cite{Mandelbrot1983}.
Time series are the most common fractal objects in most of the scientific research fields.
Lots of techniques have been proposed to analyze the fractal and multifractal properties 
of time series\cite{Peng1994,Kantelhardt2001,Kantelhardt2002,Zhou2008,Podobnik2008,Gu2010,Jiang2011}.
Multifractal detrended fluctuation analysis(MF-DFA)\cite{Kantelhardt2002} we adopted here has shown to be a quite effective
tool to investigate the multifractal properties of non-stationary time series.
Also the complex network\cite{albert2002} is one of the generic ways to describe complex systems. Transformations from
time series to networks have attracted substantial considerations recently\cite{Thiel2004,Zhang2006,Lacasa2008,Zhao2014a}.
It has been shown that the visibility graph method\cite{Lacasa2008} is suitable for characterizing the correlation and 
geometrical structure of time series\cite{Lacasa2009}.\\

Recently there has been an substantial interest in understanding how those complex 
systems behave near the critical point. Lots of early warnings have been discovered to serve as
a signals of the coming of the tipping points\cite{Scheffer2009}. In this paper, we use the MF-DFA and the visibility graph method 
to analyze the behaviour of the Ising system near critical temperature. The level of the mutifractality of the 
time series and topological quantities of the complex networks converted from time series can bee seen as some new
early warnings. Hurst exponents increase
dramatically around critical point which means the time series change from a 
weak correlated to long term correlated ones. Generalized Hurst exponents have shown the transformation of fractal  
structures of the time series from weak multifractal (or monofractal) to multifractal while the system approaches to critical regime. 
These indicate huge differences between the dynamical behaviours of Ising system at different temperatures.
Evolution of the singularity spectrum has depicted the extreme strong multifractality around critical point.
Structure parameters of the singularity spectrum suggest that the time series become much more complex around critical point.
The shuffling procedure reveals that both broad probability density function and long term correlations 
are the sources of multifractality of the magnetization time series around critical point.
Visibility graphs converted from the time series at different
temperature share the heterogeneous property which is consist with the previous results: fractal 
time series can be converted to scale-free networks\cite{Lacasa2008}. We also find that the 
increase and decrease of the topological quantities can be used to identify the coming of phase transition. 
The evolution behaviours of topological 
structures of the visibility graphs have manifested the differences among geometrical structures of
magnetization time series at different temperature regimes from network perspective.
\section*{ Methods\label{Sec:Methods}}

\subsection*{Simulation of Ising model\label{SubSec:SimuIsingModel}}

The two-dimensional Ising model is a paradigm of physical phase transition. Suppose we have a
square lattice of $N$ sites with periodic boundary conditions. A spin state $\sigma$ is defined on each site with one
of two possible orientation values which denoted by $\sigma = \pm 1$. So 
the number of all possible configurations of the system is $2^N$. The Hamiltonian is
\begin{equation}
	H(\sigma) = -\sum\limits_{<i,j>}J_{ij}\sigma_{i}\sigma_{j} -\mu\sum\limits_{i=1}^{N}h_{i}\sigma_{i}\qquad.
\end{equation}
Any two adjacent sites $i,j\in N$ have an interaction strength $J_{ij}$. A site
$i\in N$ has an external magnetic field $h_i$ acting on it and $\mu$ is the magnetic 
moment. Here we concern about the case while $J_{ij}=1$ and $h_i=0$. The order of 
the system can be measured through the magnetization per spin which defined as
\begin{equation}
	M=\frac{1}{N}\sum\limits_{i=1}^{N}\sigma_{i} \qquad.
\end{equation}
It is well known that for two-dimensional Ising model the system goes through a phase
transition when temperature $T$ is equal to critical temperature $T_c$. There exists
spontaneous magnetization as all the spins tend to equal towards either +1 state or -1
state at a non-zero critical temperature $T_c$.\\

Here in this paper we focus on the dynamics of the Ising model. 
Monte Carlo simulation by using Metropolis algorithm\cite{Metropolis1953} is employed to capture the 
evolution of the system. Thus the time series of magnetization $M$ can be 
obtained from simulation procedure.\\

In the Metropolis algorithm, new configurations are generated from the previous
state using a transition probability. The probability of the system being in a state $n$ follows the Boltzmann distribution:\\
\begin{equation}
	P_{n}=\frac{e^{-E_{n}/kT}}{Z}
\end{equation}
,where $E_n$ is the free energy of the state, $k$ is the Boltzmann constant, $T$ is
temperature and $Z$ is the partition function. Thus the transition probability from
state $n$ to $m$ is given by
\begin{equation}
	P_{n\rightarrow m}=\mathrm{exp}[-\Delta E/kT] 
\end{equation}
,here $\Delta E=E_m-E_n$.\\

According to previous description, given an initial state of the system, the 
algorithm proceeds as follows:\\
1. Choose a site $i$ randomly;\\
2. Calculate the energy change $\Delta E$ if spin site $i$ were to be flipped;\\
3. If $\Delta E$ is negative, flip of the spin of site $i$ is accepted. If
$\Delta E$ is positive, a random number is drawn between 0 and 1 uniformly and the
flip is accepted only if the random number is smaller than $\mathrm{exp}[-\Delta E/kT]$;\\
4. Choose another site and repeat the previous steps.\\

A Monte Carlo step is completed when every spin of the system has had a chance 
to flip. In the ordinary simulation schematic, due to the phenomena of critical slowing down, 
when the system is around critical 
temperature $T_c$, one should skip several Monte Carlo steps in order to avoid 
correlations between successive configurations. This is important for evaluating the 
interested quantities accurately. On the contrary, here we keep all the time series of 
magnetization $M$ calculated from every simulation step. We are indeed very 
interested about the correlations and the properties caused by the critical slowing down phenomena.\\

We set $k=1$ and $J_{ij}=1$ which means the exact critical temperature $T_c\simeq 2.27$. 
We have simulated the system from $T=1.17$ to $T=3.62$ with 
$\Delta T=0.05$. For every discrete temperature we run an ensemble of 100
simulations of 100000 Monte Carlo steps(the first 10000 steps have been discarded to overcome the 
influence of the initial configuration). In order to investigate the finite size effect, we have simulated different lattice sizes for 
$N=100\times 100,150\times 150,200\times 200,300\times 300$.\\

By using the Metropolis algorithm, we have obtained a time series ensemble at different temperature.
Then in the following section we will introduce two methods which used to 
analyze the properties of those time series from two different aspects: multifractal and complex network.
\subsection*{Mutifractal Detrended Fluctuation Analysis\label{SubSec:MF-DFA}}
We adopt multifractal detrended fluctuation analysis (MF-DFA) to analyze the hierarchy 
of scaling exponents of the magnetization time series corresponding to different 
scaling behaviour\cite{Kantelhardt2002}. It is the generalization of detrended fluctuation analysis (DFA)\cite{Peng1994}.
The MF-DFA method have been widely 
applied to characterize the properties of various non-stationary time series from 
different fields such as financial market
\cite{Jiang2008a,Ruan2011,KRISTOUFEK2012,Morales2012,Oh2012,Siokis2013,Morales2013,Hasan2015,Stosic2015,Stosic2015b},
physiological\cite{Dutta2010}, biology\cite{Duarte-Neto2014},traffic jamming\cite{Andjelkovi2015}, 
geophysics\cite{Telesca2006} and neuroscience\cite{Zorick2013}.\\

The MF-DFA method is proceeded as follows: 
$(i)$ Suppose we have a time series
$\{x(i)\},i=1,\ldots,l$. We first integrate the time series to get the profile
$y(k)=\sum\limits_{i=1}^{k}[x(i)-\langle x\rangle],k=1,\ldots,l$, where $\langle x\rangle$ is the
mean value of $\{x(i)\}$. 
$(ii)$ Divide the integrated series $y(k)$ into $l_{s}=int(l/s)$ 
non-overlapping segments of length $s$. 
Calculate the local trends for each of $l_{s}$ segments by a least-square fit
and subtract it from $y(k)$ to detrend the integrated series.
We get the detrended variance of each segment $v$\\
\begin{equation}
	F^2(v,s)=\frac{1}{s}\sum\limits_{i=1}^{s}\{y(v-1+i)-\widetilde{y(v,i)}\}^{2}\qquad
\end{equation}
,here $\widetilde{y(v,i)}$ is the fitting polynomial in segment $v=1,\ldots,l_{s}$. 
We use the three order polynomial fit here. 
$(iii)$ The step $(ii)$ has been proceed from both the beginning 
and the end of the time series which leads to $2l_s$ segments in total. Average 
over all segments to obtain the $q\mathrm{th}$ order fluctuation function\\
\begin{equation}
	F_q(s)=\{\frac{1}{2l_s}\sum\limits_{v=1}^{2l_s}[F^2(v,s)]^{q/2}\}^{1/q}\qquad.
	\label{fluctuationFunction}
\end{equation}
$(iv)$ Repeat this calculation to get the fluctuation function $F_q(s)$ for different box size $s$. 
If $F_q(s)$  increases by a power law $F_q(s)\sim s^{h(q)}$,
the scaling exponents $h(q)$ (called generalized Hurst exponent) can estimated as the slope of the linear regression of $\mathrm{log}F_q(s)$ versus $\mathrm{log}s$. $h(q)$ is 
the fluctuation parameter and describe the correlation structures of the time series at different magnitude.
The value of $h(0)$ can not be determined by using Eq(\ref{fluctuationFunction}) because of the diverging 
exponent. The logarithm averaging procedure should be used,\\
\begin{equation}
	F_{0}(s)=\mathrm{exp}\{\frac{1}{4l_s}\sum\limits_{v=1}^{2l_s}\mathrm{ln}[F^{2}(v,s)]\}\sim s^{h(0)}\qquad.
\end{equation}
Generalized Hurst exponents $h(q)$ as a function of the $q\mathrm{th}$ order can quantify the multifractality.
If $h(q)$ are the same for all $q$, the time series is monofractal or the time series is mutifractal.\\

The classical multifractal scaling exponents\cite{Barabsi1991}\\
\begin{equation}
	\tau(q)=qh(q)-1
	\label{classicalExponent}
\end{equation}
are directly related to the generalized Hurst exponents $h(q)$.
Another way to characterize multifractal time series is by
using the singularity spectrum $f(\alpha)$ which is related to 
$\tau(q)$ via the Legendre transform\cite{Kantelhardt2002},\\
\begin{equation}
	\alpha=\tau^{'}(q),f(\alpha)=q\alpha -\tau(q)\qquad.
\end{equation}
Using Eq(\ref{classicalExponent}), we have\\
\begin{equation}
	\alpha=h(q)+qh^{'}(q),f(\alpha)=q[\alpha-h(q)]+1\qquad.
	\label{singularitySpectrum}
\end{equation}
Here $\alpha$ is the singularity strength. Geometrical shape of the singularity 
spectrum can illustrate the level of mutifractality from 
three parameters: position of the maximum $\alpha_0$ where $f(\alpha)$ reaches its maxima; 
width of the spectrum $W=\alpha_{max}-\alpha_{min}$ which
can be obtained from extrapolating the fitted $f(\alpha)$ curve to zero; 
and skew shape of the spectrum $r=(\alpha_{max}-\alpha_0)/(\alpha_0 - \alpha_{min})$.
In a nutshell, small value of $\alpha_0$ means the underlying process is more regular. 
The wider the spectrum (larger $W$),
the richer the structure of time series.
Right skew shape of the spectrum with $r>1$ indicates more complex than left skew shape with $r<1$\cite{SHIMIZU2002,Stosic2015}.	
\subsection*{Visibility Graph Method\label{SubSec:VGM}}
Complex network and time series are two generic ways to
describe complex systems. Dynamical properties of time
series can usually be preserved in network topological
structures. Lots of methods have been developed to capture the geometrical 
structure of time series from complex network aspect such as 
cycle network\cite{Zhang2006}, correlation network\cite{Yang2008}, 
visibility graph\cite{Lacasa2008}, recurrence network\cite{Xu2008}, isometric
network\cite{Zhao2014a} and many others. It has been reported that the visibility graph can
capture the geometric properties of time series by network
structure and it has a straight-forward geometric interpretation of the original time series.
That is periodic time series can be transformed into regular
networks and random series corresponding to random networks\cite{Newman2002c}. Moreover, fractal
series can be converted into scale-free networks\cite{Barabasi1999}. 
Hence the visibility graph method has been successfully applied 
in many fields\cite{Yang2009,Liu2010,Andjelkovi2015}\\
\indent  The visibility graph procedure can be described as
follows: For a time series $\{x(i)\},i=1,\ldots,l$, the final 
transformed network has $l$ vertices. Two arbitrary
data points $(t_i,x(i))$ and $(t_j,x(j))$ will have visibility, 
and those two data points will become two connected nodes $i$ and $j$ 
of the associated network connected by edge $e_{ij}$, if all the
data points $(t_k,x(k))$ placed between them fulfil:\\
\begin{equation}
	x(k) < x(j) + (x(i)-x(j))\frac{t_j - t_k}{t_j - t_i}.
\end{equation}
The network obtained from this algorithm will always be connected. 
The network is also undirected because we do not keep any
direction information in the transformation.

\section*{Results and Discussion\label{Sec:Results}}
\subsection*{Statistical Properties of Time Series\label{SubSec:SPTS}}
\begin{figure}[!ht]
	\centering
	\includegraphics[width=\linewidth]{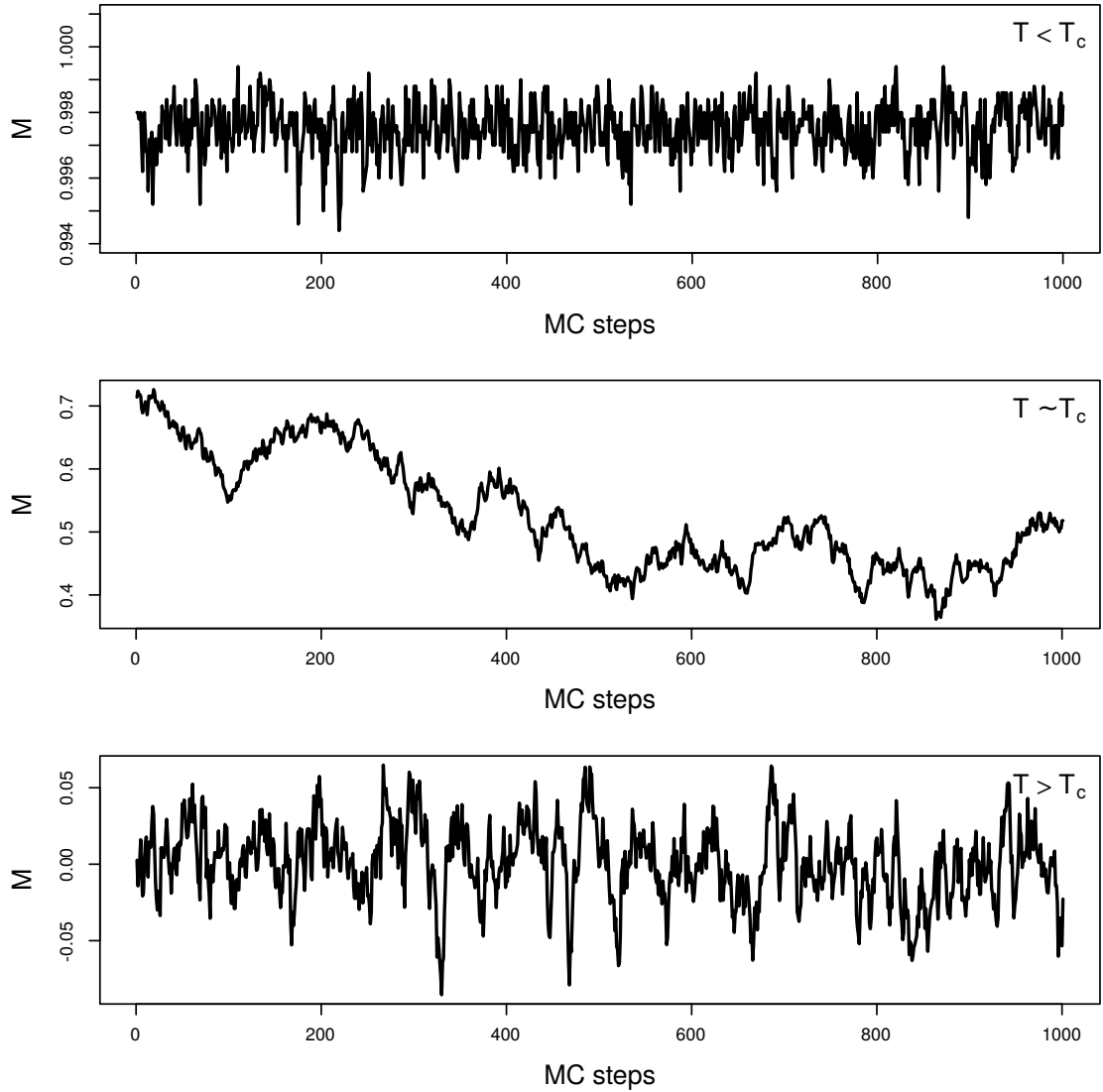}
	\caption{{\bf Magnetization per spin time series.} Magnetization as a function of time in the two-dimensional 
		Ising model for different temperatures.}
	\label{fig1}
\end{figure}
Fig.~\ref{fig1} shows the time series at three regimes $T<T_c$, $T\simeq T_c$, and $T>T_c$ from
top to bottom respectively. Obviously we can observe different outlines of the time
series at different temperature regimes. Detailed properties of those time
series will be analyzed by using multifractal detrended analysis and visibility graph method in the following context.\\
\begin{figure}[!ht]
	\centering
	\includegraphics[width=\linewidth]{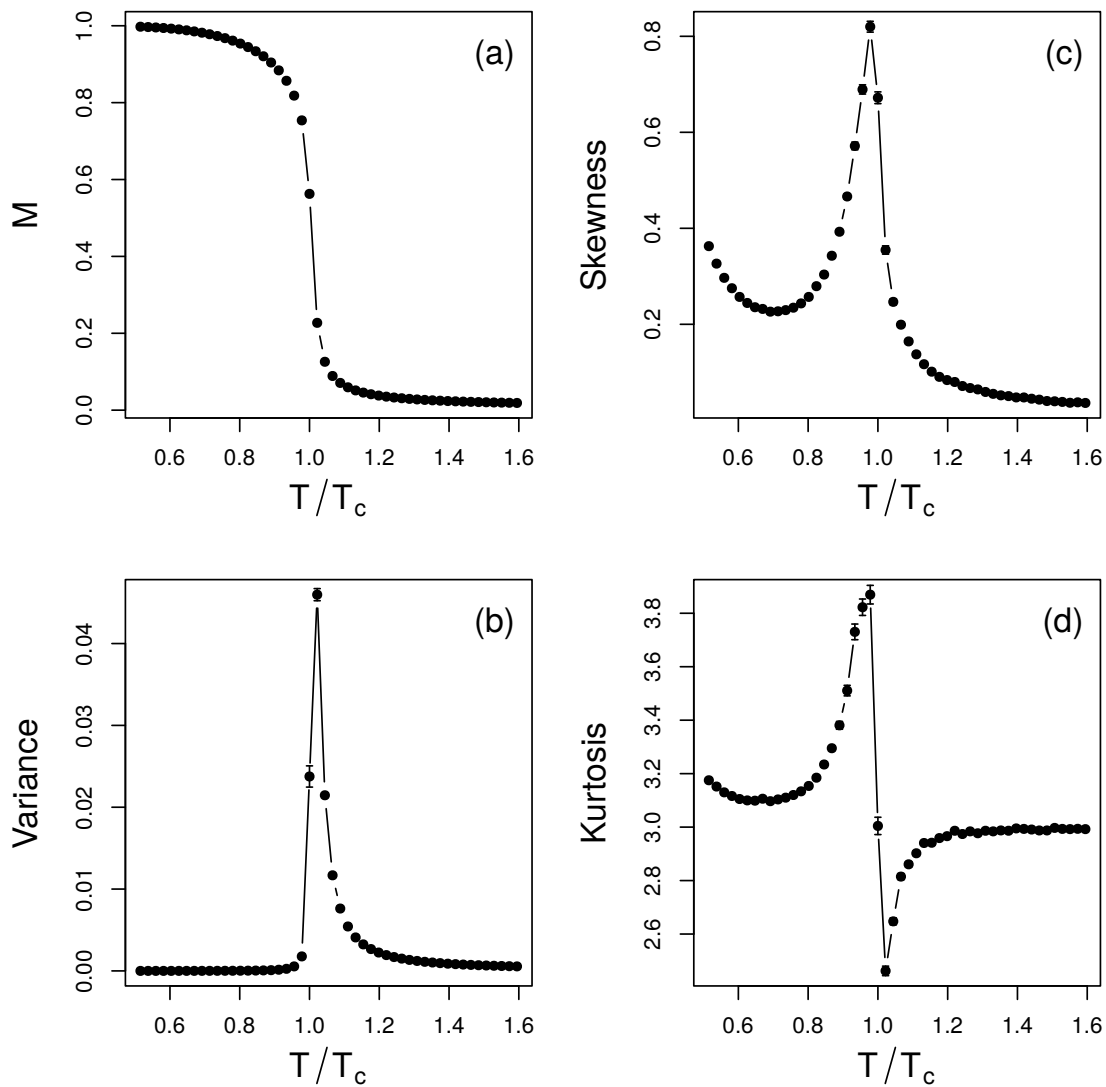}
	\caption{{\bf First four moments of the magnetization time series.} Ensemble average of (a) mean, (b) variance,
		(c) skewness (d) kurtosis of the magnetization time series versus the relative temperature, $T/T_c$ for system size $N=100\times 100$.}
	\label{fig2}
\end{figure}

The trends of magnetization time series at different temperature are obviously different.
We use first four order moments to quantify the distribution differences between those time series
in Fig.~\ref{fig2}. The ensemble standard errors are defined as standard deviation divided by
square root of ensemble size (100) and are shown with error bars in figures. 
Fig.~\ref{fig2} (a) is the ensemble average magnetization 
vary with relative temperature $T/T_c$.
Simulation results have confirmed the existence of phase transition around theoretical
critical temperature $T_c\simeq 2.27$. This transition can also be verified from
three statistical quantities in Fig.~\ref{fig2} :
(b) variance, (c) skewness and (d) kurtosis which have been fully discussed in Ref\cite{Morales2015}.
Those moments have been used as early warnings of phase transition therein.
The variance is almost zero at non-critical regimes, but it becomes relatively large
around critical point. The abrupt change at low temperature regime is obviously different from
the continuous evolution at high temperature regime. Thus the variance can not be recognized as 
a useful early warning when reaching the system from low temperature regime.
Skewness is related to the asymmetry of events in the time series.
It is large only in the low temperature regime and reaches its 
maxima at critical point and it becomes very small at high temperature regime.
This is due to the meta-stable states at low temperature\cite{Morales2015}. The stochastic permutation is not strong
enough to make the system escape from the meta-stable states. This makes the distribution of 
magnetization asymmetric. The kurtosis fluctuates severely around critical point and becomes very
close to the reference normal distribution (which has a kurtosis equal to 3) when $T\ll T_c$
and  $T\gg T_c$. Apparent distribution differences between magnetization time series at different 
temperature regimes give a hint about the structure heterogeneous which demands more detailed investigations.
\subsection*{MF-DFA\label{SubSec:MF-DFAresult}}
\begin{figure}[!ht]
	\centering
	\includegraphics[width=\linewidth]{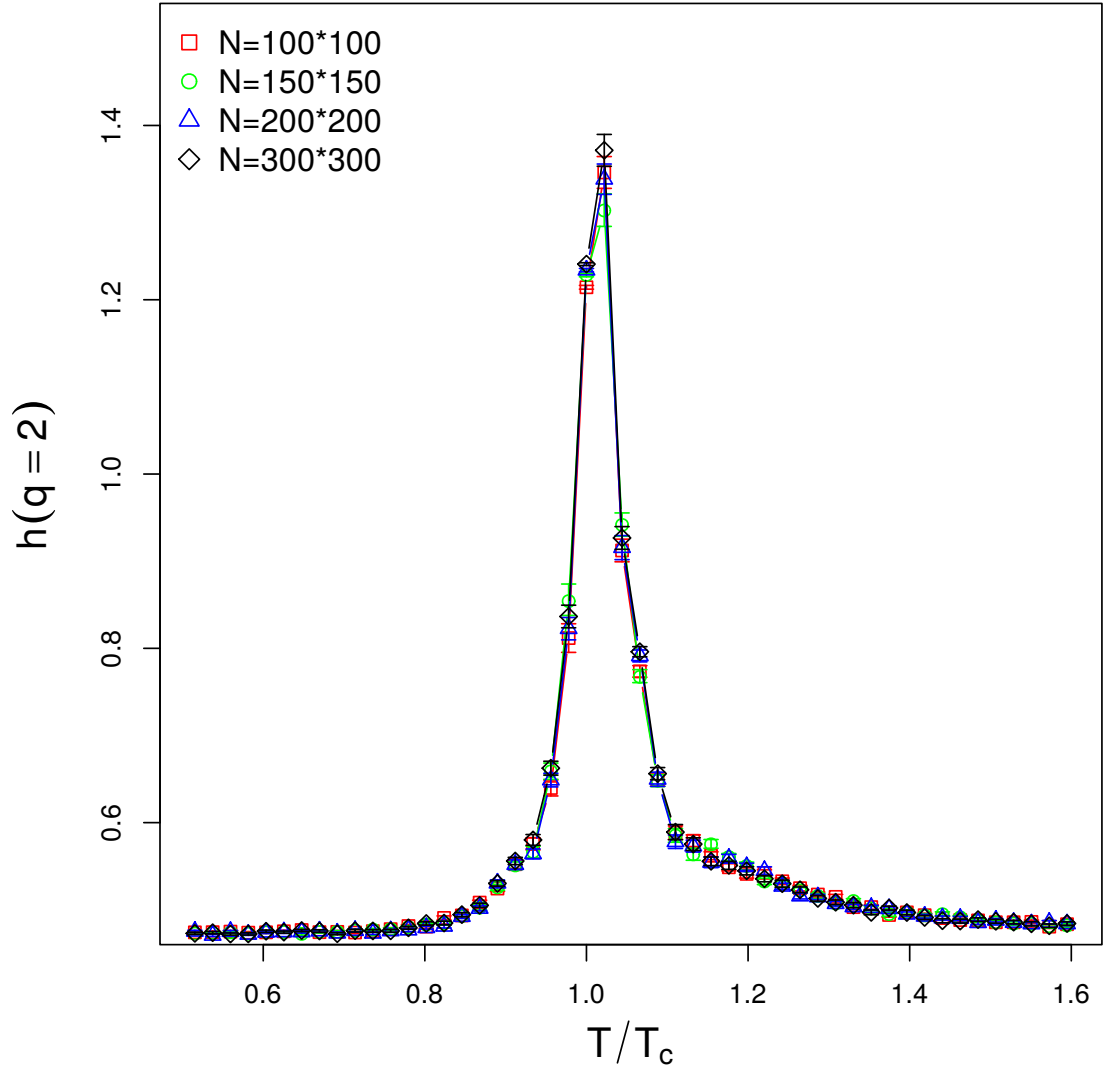}
	\caption{{\bf Hurst exponent at different temperature.} (Color online) Ensemble average of the Hurst exponent $h(q=2)$ versus the relative temperature $T/T_c$ for different system sizes.}
	\label{fig3}
\end{figure}

If we set $q=2$ in the MF-DFA, we get the same results as standard DFA method. The Hurst
exponents $h(q=2)$ have been used as early warning to detect the rising memory in time series of a system
close to critical point\cite{Landa2011,Dakos2012}.
We have calculated the ensemble average of Hurst exponent $h(q=2)$ in Fig.~\ref{fig3}. 
The Hurst exponent $h(q=2)\simeq 0.5$ when $T\ll T_c$ and $T\gg T_c$ which means 
the magnetization time series are short range correlated at these two regimes. 
When $T\sim T_c$ the Hurst exponent increases rapidly which results in $h(q=2)>1$. This means
the time series becomes non-stationary which turns to be a
unbounded process. The Hurst exponent $h(q=2)\geq 1.2$
when temperature gets very close to critical temperature. This indicates that the magnetization time
series posses long range correlation when the system approaches critical regime. The 
Hurst exponent behave almost the same when we use different system size.\\
\begin{figure}[!ht]
	\centering
	\includegraphics[width=\linewidth]{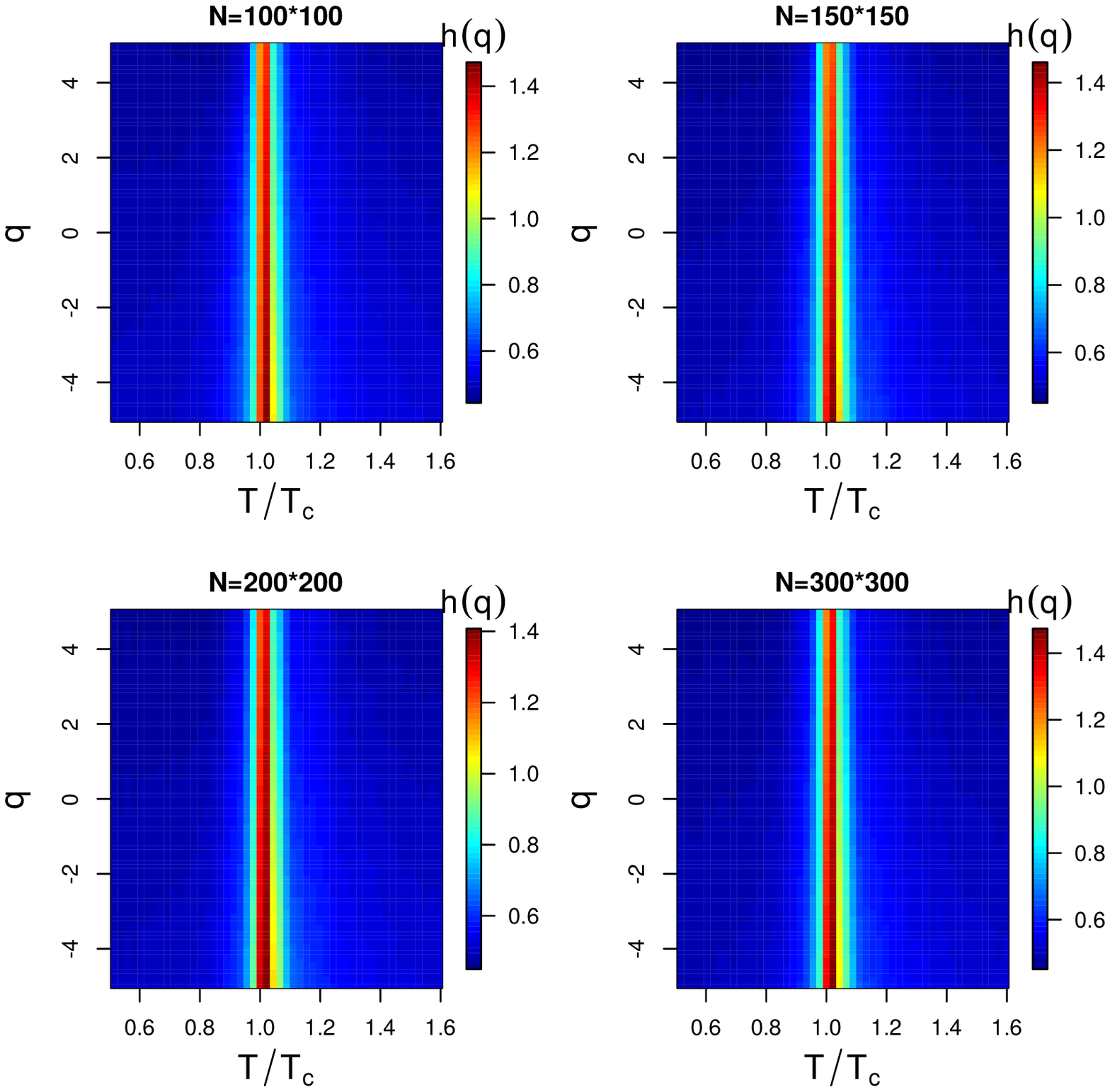}
	\caption{{\bf Generalized Hurst exponent at different temperature.} (Color online) Heat map of ensemble average of the generalized Hurst exponent $h(q)$ 
		for $q\in [-5,5]$ at different temperature regimes for different system sizes.}
	\label{fig4}
\end{figure}

Fig.~\ref{fig4} shows the generalized Hurst exponents $h(q)$ at different temperatures
for $q\in [-5,5]$ with $\Delta_q=0.1$. The system sizes are $N=100\times100,150\times150,200\times200,300\times300$. When
$T\ll T_c$ and $T\gg T_c$ the weak dependence on $q$ of $h(q)$ show the weak multifractal (monofractal) structures 
of time series. As the system 
approaches the critical regime, $h(q)$ become strongly depend on $q$. The small and large fluctuations 
of the Ising system near critical point display different scaling behaviours.
The strong non-linear dependence on $q$ of $h(q)$ around $T_c$ reveals the
strong mutifractal nature of the Ising system in the time domain.
Transformation of the multifractal properties uncover the apparent structural
and dynamical differences of the Ising system at different temperature regimes.\\

Heat map of $h(q)$ for different $q$ at different temperatures presented in Fig.~\ref{fig4} has given a
comprehensive description about scaling behaviours of the fluctuations at different magnitude. Dramatic increase of the generalized
Hurst exponent around critical temperature at different order $q$, not only at $q = 2$, can be observed.
We can see that the generalized Hurst exponent is very close to 0.5 and then it
becomes remarkably larger than 1 around critical point at all observed order $q \in [-5,5]$.
This indicates that the generalized Hurst exponent is also a very good indicator of phase transition.
Multifractality of the time series increase remarkably around critical point. Those characteristics should also be 
very good early warnings\cite{Morales2015,Scheffer2009}.\\
\begin{figure}[!ht]
	\centering
	\includegraphics[width=\linewidth]{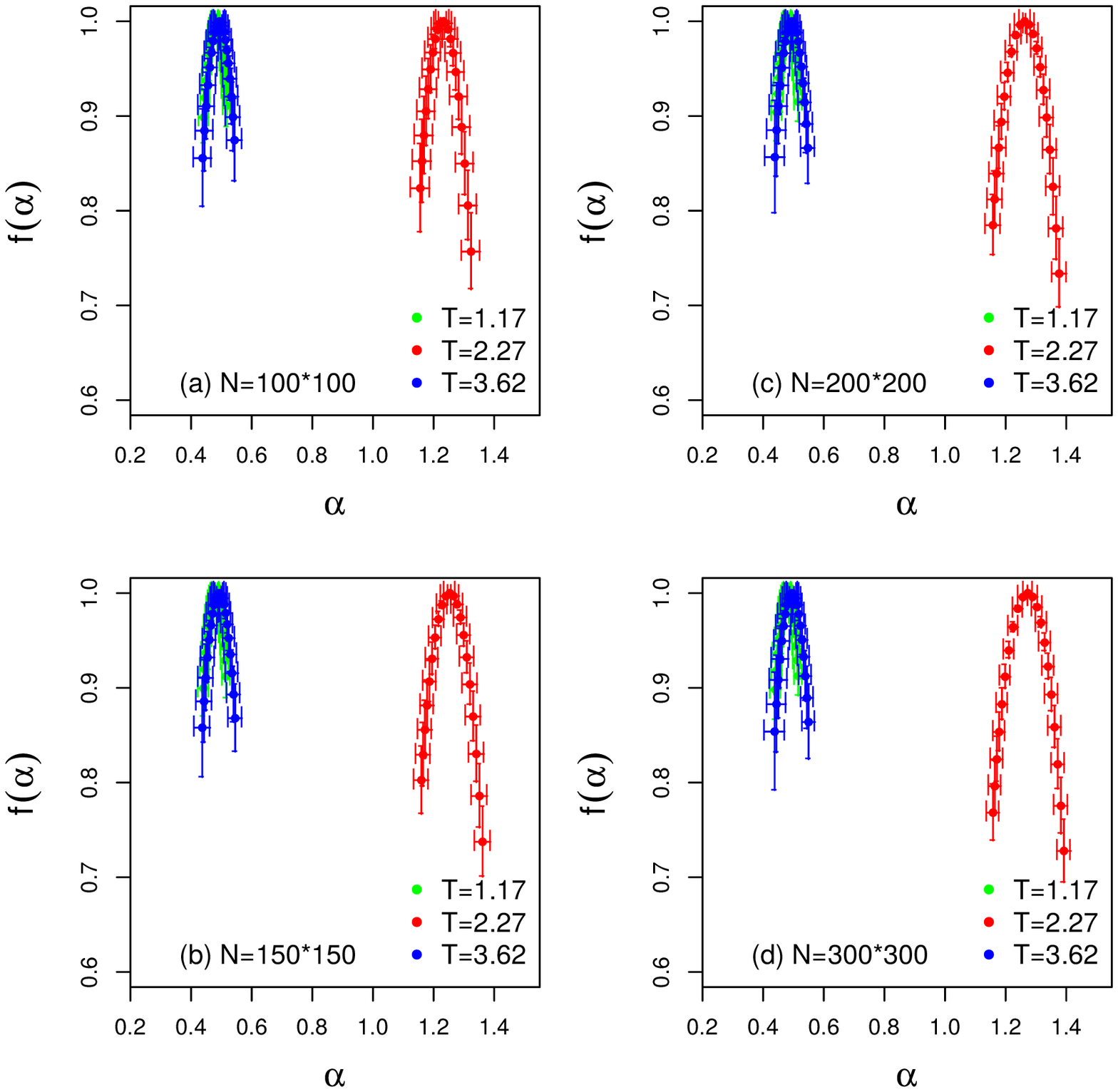}
	\caption{{\bf Singularity spectrum at different temperature regimes.} (Color online) Singularity spectrum $f(\alpha)$ of the time series as a function of the 
		singularity strength $\alpha$ at different temperature regimes for different system sizes. }
	\label{fig5}
\end{figure}

In order to quantify the level of mulfitracality, the singularity spectrum of the time series at different temperatures for different system sizes
have been presented in Fig.~\ref{fig5}. The maximum 
of the $\alpha_0$ increases to 1.3 when $T\sim T_c(2.27)$, but it decreases to 0.5 when temperature 
deviates form $T_c$. 
We can observe that the width of singularity spectrum increases as $T$ approaches $T_c$.
At the same instant the singularity becomes right skew. 
Those changes indicate more complex time series around critical temperature.
The evolution of the singularity spectrum at different temperatures are basic consistent for different system sizes.\\

A more quantitative measure about the level of complexity of the series can be given by fitting the singularity
spectrum\cite{Stosic2015} and calculating the multifractal spectrum parameters: position of maximum $\alpha_0$; width 
of the spectrum $W=\alpha_{max} - \alpha_{min}$; and the skew parameter $r=(\alpha_{max} - \alpha_0)/(\alpha_0 -\alpha_{min})$ as 
described in previous method section. A large value of $\alpha_0$ suggests the time series is irregular. Larger value of $W$ means
 richer structure of the series. Meanwhile the skew parameter $r$ determines which fractal exponents are dominant.
These parameters lead to a comprehensive measure of the singularity complexity of the time series: a series with large value of $\alpha_0$,
 a wide range $W$ of fractal exponents, and  a right-skewed shape may be considered more complex than one with opposite characteristics\cite{SHIMIZU2002}.\\
\begin{figure}[!ht]
	\centering
	\includegraphics[width=\linewidth]{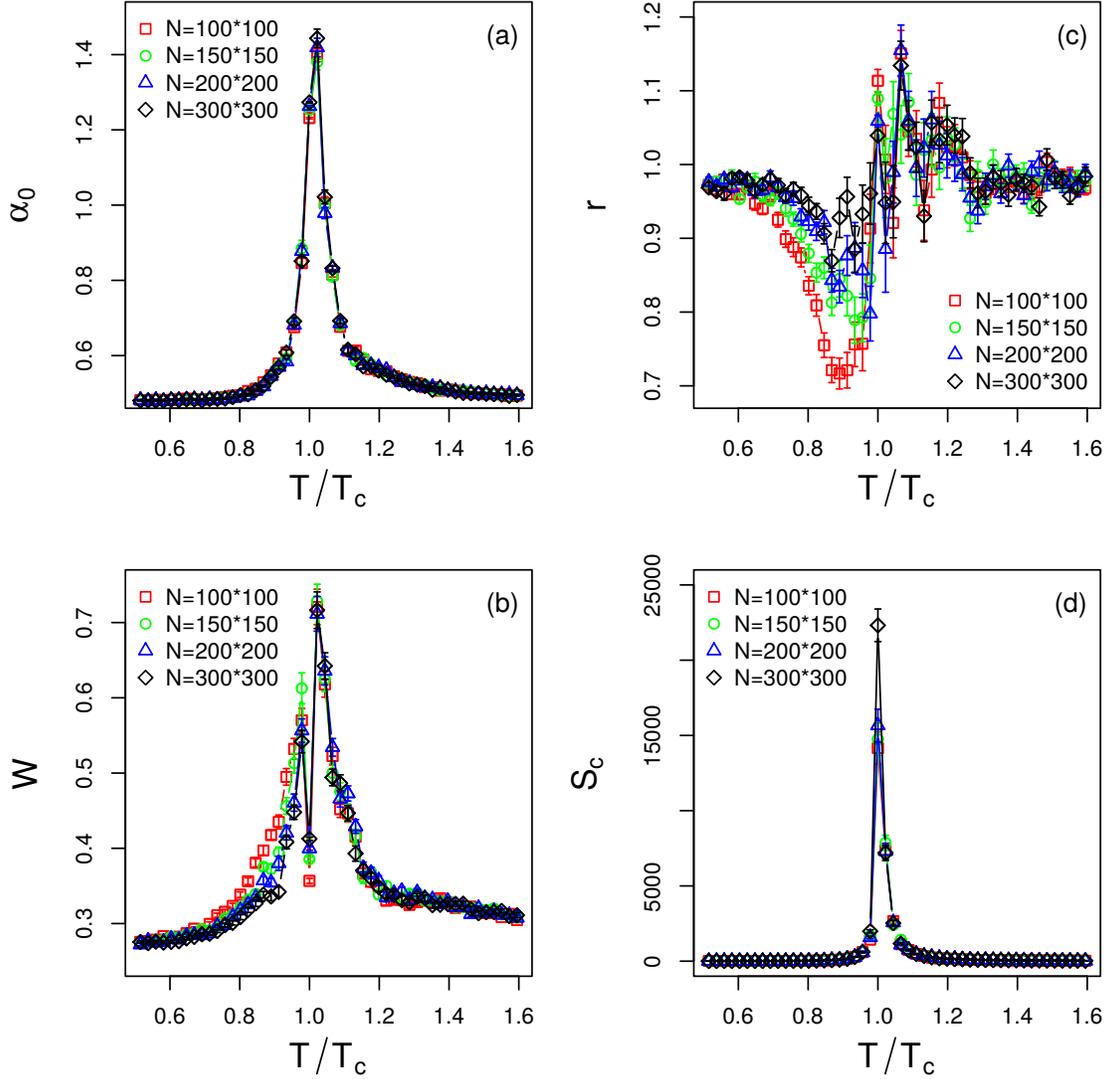}
	\caption{{\bf Complexity measure of the singularity spectrum and autocorrelation length of time series.} (Color online) The complexity measure of the time series at different temperature: (a) the position of maximum $\alpha_0$, 
		(b) the width of the spectrum $W$, (c) the skew parameter $r$ for different system size, (d) critical length 
		of the autocorrelation length $S_c$ for different system sizes.}
	\label{fig6}
\end{figure}

As shown in Fig.~\ref{fig6} (a) the value of $\alpha_0$ becomes very large at $T\sim T_c$ which suggests that the series become
extremely irregular. The evolution pattern of $\alpha_0$ can serve as a very good early warning about the coming of the critical transition.
The increase of the width $W$ of the spectrum in Fig.~\ref{fig6} (b) indicates richer structure near critical regime.
The abrupt jump right at the critical point gives a hint about the mutation of the correlation structure. 
The skew parameters $r$ in Fig.~\ref{fig6} (c) are almost equal to 1 when temperature deviates from critical regime which manifest symmetry shapes of the multifractal spectrum
at low and high temperature regimes. It becomes larger than 1 at critical threshold. 
One interesting finding is that the skew parameter $r$ first gradually decreases at lower temperature regime and it increases
very fast when the system gets particularly close to critical temperature. This shows that the large fluctuations dominate when
the system approaches critical regime from the lower temperature. On the contrary the skew parameter $r$ will slowly increase 
to the maxima when the system moving close to $T_c$ from the right side. Thus the small fluctuations contribute the most to 
the multifractality at critical point. This indicates a more complex structure with right skewed shape above the critical regime.\\
\begin{figure}[!ht]
	\centering
	\includegraphics[width=\linewidth]{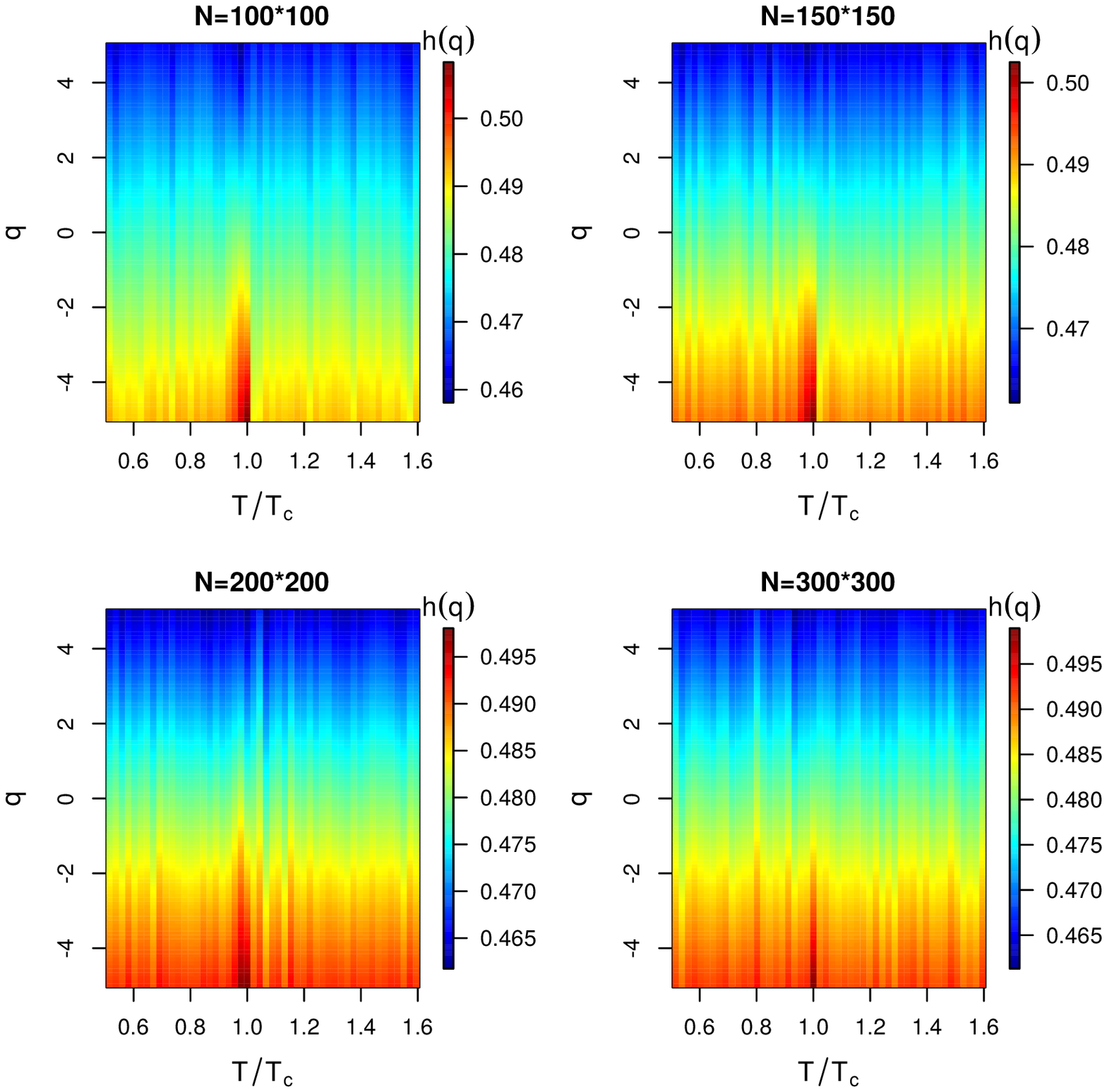}
	\caption{{\bf Generalized Hurst exponent for shuffled time series.} (Color online) Heat map of ensemble average of the generalized Hurst exponents $h(q)$ for $q\in [-5,5]$ 
		at different temperature regimes for the shuffled time series.}
	\label{fig7}
\end{figure}

\begin{figure}[!ht]
	\centering
	\includegraphics[width=\linewidth]{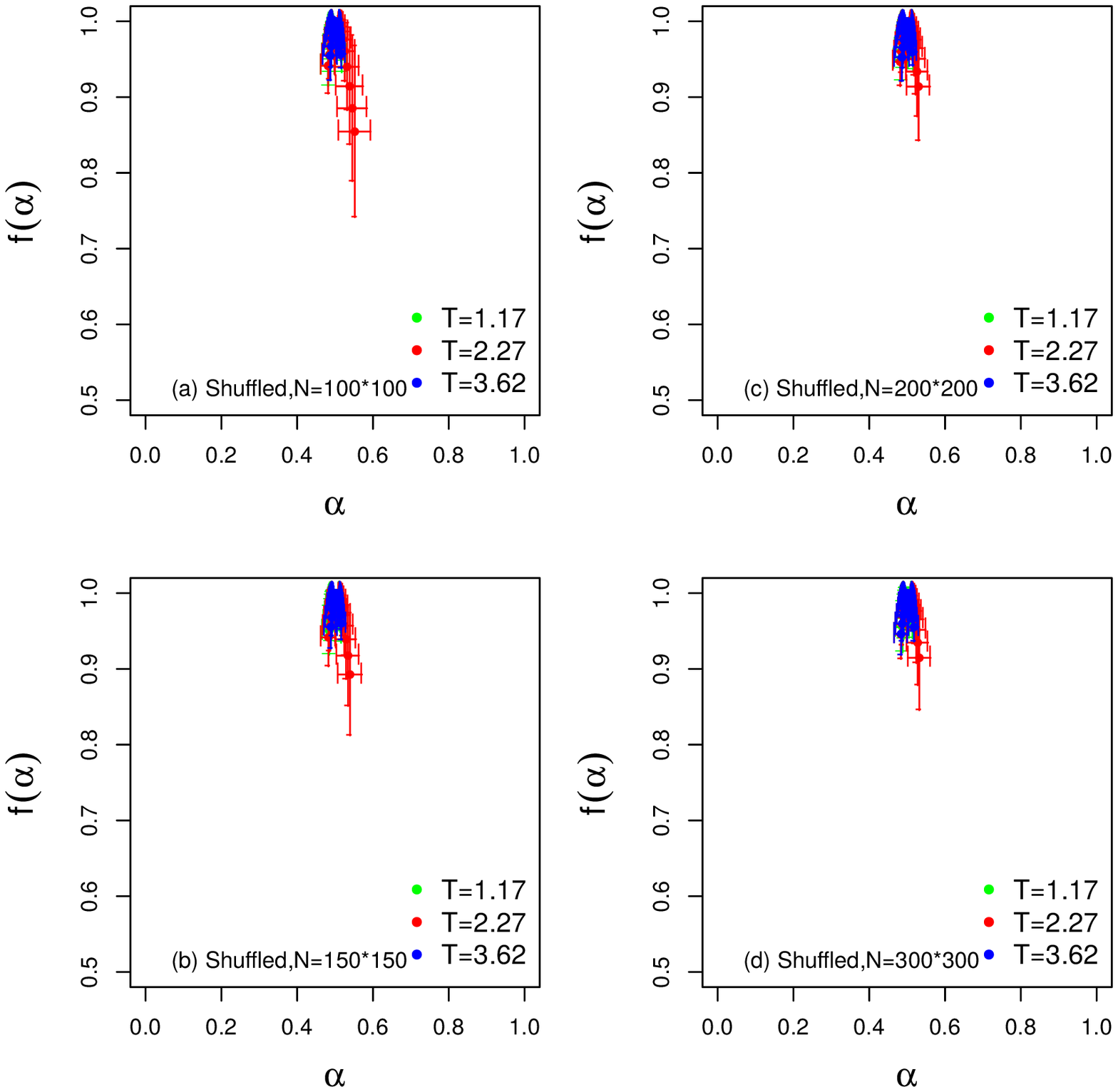}
	\caption{{\bf Singularity spectrum for shuffled time series.} (Color online) Singularity spectrum of the original time series and the singularity spectrum of the shuffled time series.}
	\label{fig8}
\end{figure}

There are two main sources of multifracality which we would like to distinguish\cite{Kantelhardt2008,Zhou2012a}: (i)
Multifractality due to broad probability density function. (ii) Multifractality
due to different long-term correlations of small and large fluctuations. Fig.~\ref{fig6} (d) gives the
critical length of the autocorrelation as a function of relative temperature. The critical length of the autocorrelation 
function is the maximum lag at which the autocorrelation smaller than the critical value $2/\sqrt{l}$. According to a 
recent research about the DFA on auto regression process (AR(1)), the autocorrelation length can be used to help estimating the Hurst 
exponent more accurately\cite{Marc2015}. It is known that the
two point correlation function of Ising system will become divergence in the thermodynamic limit (large N) at critical point.
The divergence of two point correlation makes the autocorrelation of the magnetization time series divergent.
Thus the strong multifractality of the Ising system around critical point should at least caused in part by the divergence of autocorrelation. 

Here we proceed the same analysis as in Fig.~\ref{fig4} and \ref{fig5}, but shuffling the data randomly to better
identify the source of multifractality. Multifractality caused by broad probability density function can not 
be fully eliminated by the shuffling procedure. Meanwhile multifractality induced by long-term correlations
can be removed after shuffling the time series. The shuffling procedures have been performed $1000\times l$ transpositions on each series
with 100 ensemble average. While $l=100000$ is the length of each time series.
Fig.~\ref{fig7} gives the dependency between $h(q)$ and $q$ for the shuffled
time series at different temperature under different system sizes. The strong non-linear dependency only exist when $T\sim T_c$.
In Fig.~\ref{fig8} we show the singularity spectrum of shuffled time series at
three temperature regimes for different system sizes. We find that the singularity spectrum at $T=1.17$ and $T=3.62$ are nearly at one point
with $\alpha_0 \simeq 0.5$. On the contrary the singularity spectrum of the time series at $T=2.27$ 
still posses multifractality with a right skew shape. We address that the source of strong multifractality of 
the system near critical regime stems from both long-term correlations and  broad probability density function\cite{Barunik2012}.\\
\subsection*{Visibility Graph\label{SubSec:VGrsults}}
We use the visibility graph method to convert magnetization 
time series under different temperatures to complex networks.  
The networks' sizes are 10000 which converted from time series 
with length 10000. Then different topological quantities have been calculated for 
networks obtained at different temperatures. Those topology quantities\cite{albert2002}
can characterize the geometrical properties of time series which 
are directly related to the dynamics of Ising system.\\

\begin{figure}[!ht]
	\centering
	\includegraphics[width=\linewidth]{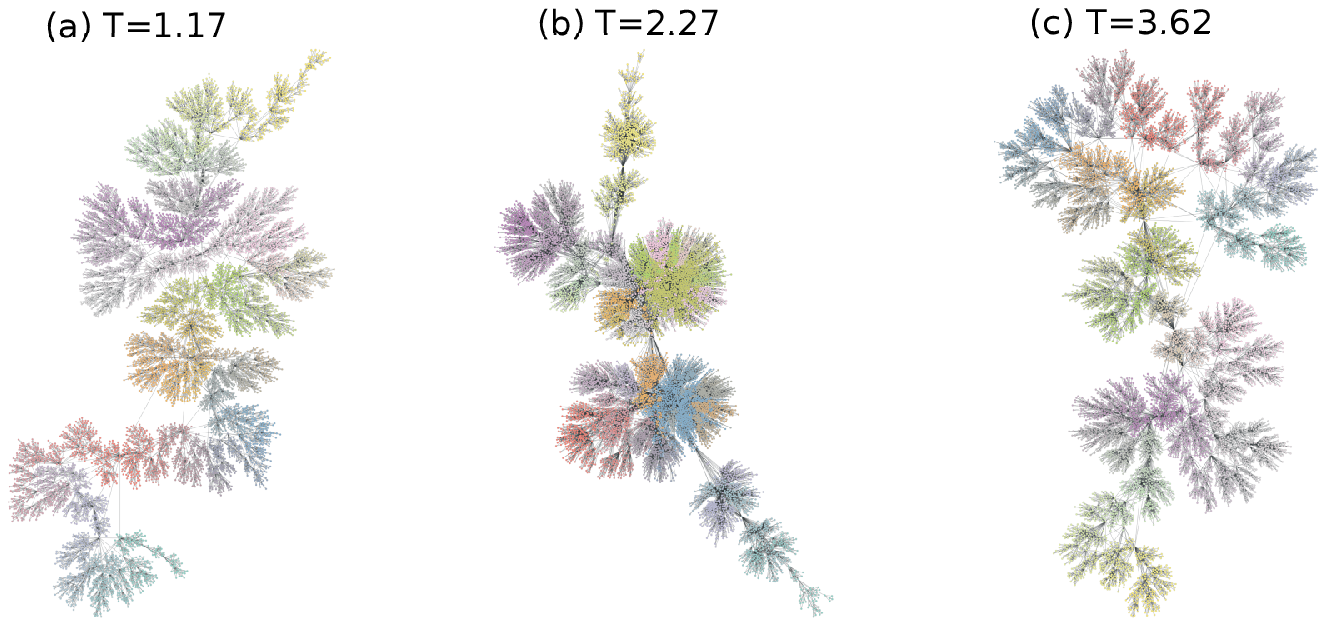}
	\caption{{\bf Visibility graphs at different temperature regimes.} (Color online) The network structure of the visibility graphs at (a) $T=1.17$, (b) $T=2.27$, (c) $T=3.62$. The network sizes are 10000. Different colors represent
		different communities. }
	\label{fig9}
\end{figure}

Here we give three networks obtained via visibility graph method from three different time series at three different temperatures in 
Fig.~\ref{fig9}. Fig.~\ref{fig9}(a), (b), (c) are the visibility graphs at $T=1.17, 2.27, 3.62$ respectively. The network at $T=2.27$
is markedly different from two networks at $T= 1.17, 3.62$. The extremely modular network structure is exhibited in Fig.~\ref{fig9}(b).
It can be understood that when $T=2.27$ the large trends of the time series make some extreme values have massive visibilities.
This is responsible for the formation of large communities which presented by
different colors. Two networks in Fig.\ref{fig9}(a) and (c) possess tree like 
structures. The community sizes of these two networks are small than the network when $T=2.27$.
The lack of trends leads to the snowflake shapes. Thus the geometric outlines of the time series
at different temperature regimes are preserved due to the affine invariant features of the visibility graph method.
\begin{figure}[!ht]
	\centering
	\includegraphics[width=\linewidth]{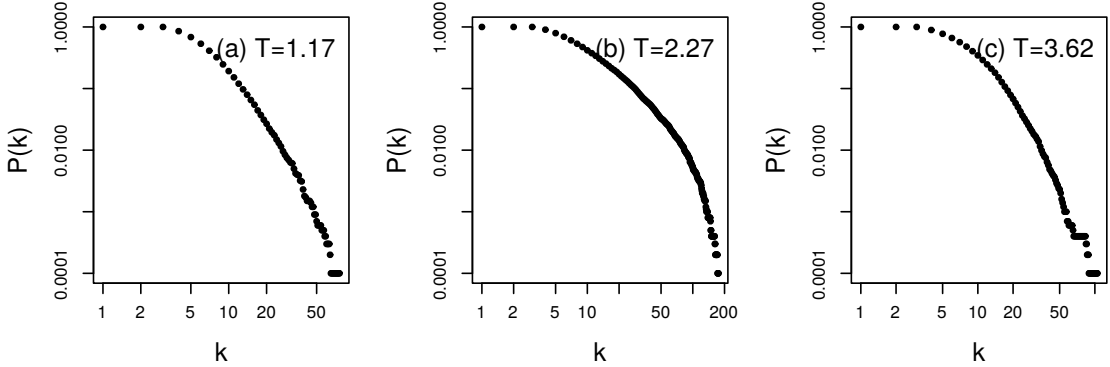}
	\caption{{\bf Cumulative degree distributions of visibility graphs.} Cumulative degree distributions of networks at three different 
		temperature regimes with (a) $T=1.17$, (b) $T=2.27(T_c)$, (c) $T=3.62$ respectively for system size
		$N=100\times 100$.}
	\label{fig10}
\end{figure}

\begin{figure}[!th]
	\centering
	\includegraphics[width=\linewidth]{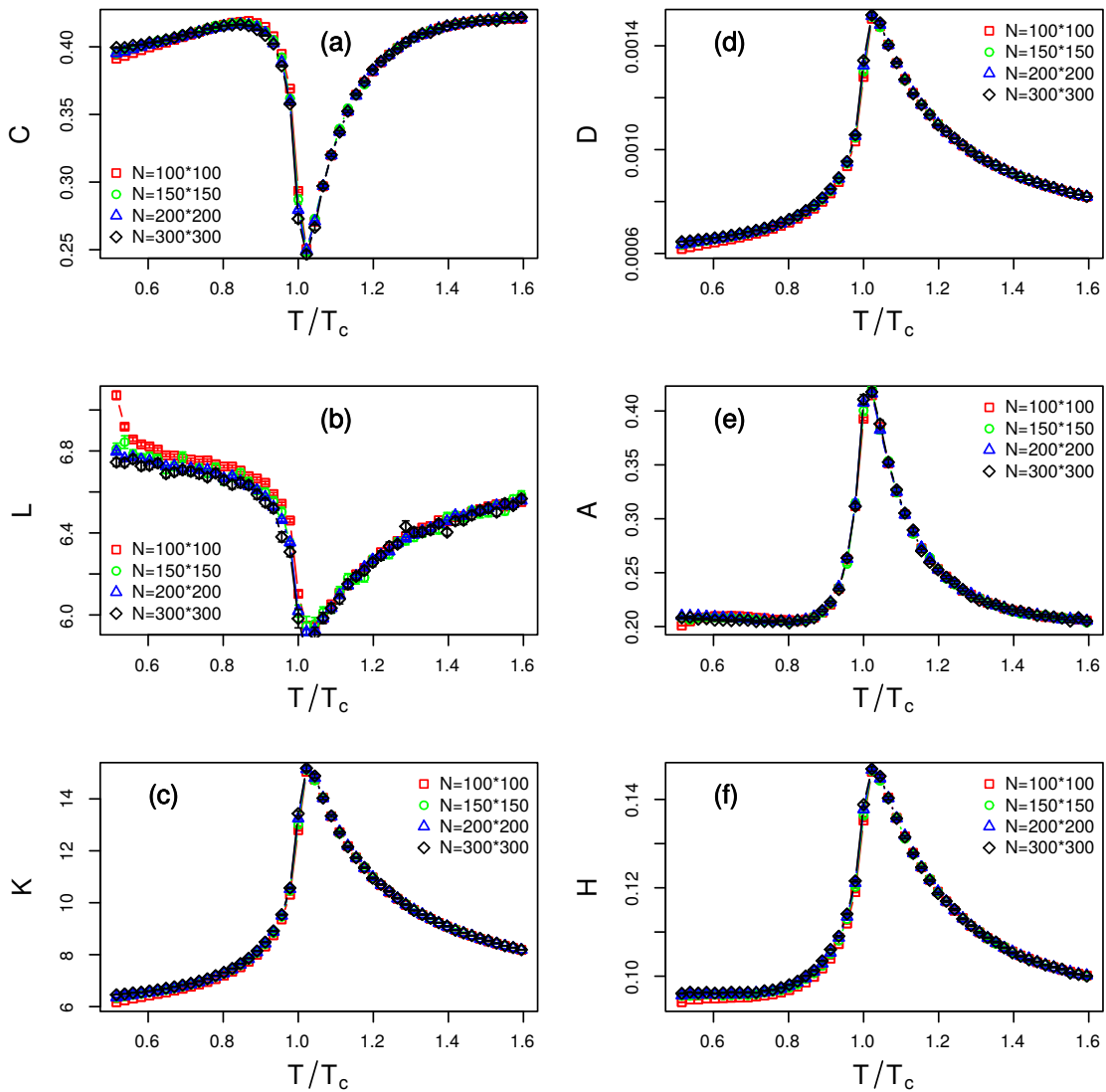}
	\caption{{\bf Topology quantities of visibility graphs.} (Color online) Ensemble average of the topology quantities (a) clustering coefficient $C$
		, (b) average degree $K$, (c) average shortest path length $L$, (d) the density $D$, 
		(e) the assortativity $A$, (f) the heterogeneity $H$ of networks converted from time series at different temperature. Different colors and symbols represent different system sizes.}
	\label{fig11}
\end{figure}

Fig.~\ref{fig10} gives the degree distributions of networks
at three temperature regimes. Fig.~ \ref{fig10} (a), (b), (c) are 
degree distributions of networks at temperature $T=1.17$, $T=2.27(T_c)$ and 
$T=3.62$ respectively.
Three distributions show the heterogeneous nature of
the networks at three temperature regimes. The network at $T=T_c$ with broader degree distribution
is apparently more heterogeneous than those at $T<T_c$ and $T>T_c$. The largest degree of network at $T=T_c$ is 288 which 
is significantly larger than those networks' largest degree away from critical temperature.
This can also be verified in Fig.~ \ref{fig11} (f): heterogeneity\cite{estrada2010quantifying} of the networks
increase dramatically near critical point. 
According to Ref\cite{Lacasa2008},
fractal time series will be converted to scale-free network. So from
the degree distributions of networks we can gain some insights about
the fractal nature of magnetization time series which has been discussed
by using MF-DFA.

\indent In Fig.~\ref{fig10}, networks at different regimes share the heterogeneous nature, but they
do have some differences which are related to the dynamic properties of Ising
system at specific temperature. In Fig.~ \ref{fig11}, we calculate the ensemble
average of topology quantities at different temperature.
In Fig.~ \ref{fig11} (a), the clustering coefficient $C$, (b) average shortest path length $L$ both decrease 
around the critical point. Decreases of $C$ and $L$ around critical point indicate more heterogeneous network structures.
We can verify these changes from Fig.~ \ref{fig11} (f). According to Ref\cite{estrada2010quantifying}, heterogeneity of the 
BA network\cite{Barabasi1999} is around 0.11 which is much smaller than the heterogeneity of network at critical point. 
The extreme heterogeneous structure is due to utmost non-stationarity and  long-term correlations of the time series 
which is exactly the characterizations of critical state of Ising system.
The average degree $K$, network density $D$ in Fig.~ \ref{fig11} (c) and (d) reach their maxima around 
critical temperature. The network becomes more and more dense when the system approaches $T_c$. Fig.~ \ref{fig11} (e)
shows the assortativity $A$ as a function of relative temperature. The fast increase of $A$ around $T_c$ gives a hint about
the coming of the critical threshold. It also tells that the network becomes more and more assortative which means the existence 
of long trends and extreme values in the time series near $T_c$.
All those topology quantities transitions suggest huge structure distinctions
between networks at different temperature regimes. Those results have
shown geometrical structure transition of the time series at different regimes
which are signals of phase transition of Ising system from the view of complex networks.
The way of those topological quantities approach the critical point either form the low temperature side or 
the high temperature side have manifest the possibility of been used as early warnings.\\

As a final remark of the paper, we shall emphasize the importance of our previous results from the view of early warnings\cite{Scheffer2009,Kefi2014,Morales2015}.
Variety of early warning signals have been proposed and summarized \cite{Scheffer2009,Dakos2012,Faranda2014,Morales2015}. 
Those early warnings can be categorized into two main classes: metric-based indicators which probing the delicate
changes in the statistical properties of the time series, and model-based indicators which detect changes in the 
time series dynamic fitted by a reasonable model. Thus our results shown in previous sections are metric-based
estimators. The four order moments given in Fig.~\ref{fig2} has been fully discussed in 
ref \cite{Morales2015}. It is exactly consistent with previous researches that the critical slowing down and
the fluctuation patterns of the complex system near critical thresholds will make the autocorrelation, variance and
skewness increase\cite{Ives1995,Guttal2008,Lenton2009}. The contribution of our work is that the MF-DFA and
visibility graph elaborate the multifractal and geometrical properties of the Ising system near critical point in the time domain.
So far as we know this is the first time to use the variation of the multifractal  and geometrical properties of the system as early
warnings. In fact due to the divergence of the spatial correlation of Ising system near critical threshold, the 
spatial multifractal features may also be used as early warning signals. We can use the higher dimension MF-DFA\cite{Gu2006,Carbone2007,Zhou2013} to
explore this problem. This should be subject to future investigations.

\section*{Conclusions\label{Conclusions}}
In conclusion, we have used the mutifractal detrended analysis (MF-DFA) and the visibility graph method to analyze the 
outputs of two-dimensional Ising model - magnetization time series. Dynamics of the system at 
different temperatures are directly related to the multifractal and geometrical properties of time series.
First four order statistical moments confirmed the existence of phase transition around theoretical critical 
temperature. The variance, skewness and kurtosis have been shown as three very efficient early warnings as
discussed in recent Ref\cite{Morales2015}. 
According to the MF-DFA, classical Hurst exponent $h(q=2)$ shows the extreme non-stationarity of 
magnetization time series at critical temperature. Also the generalized Hurst exponents uncover
the transformation of time series form weak multifractal(monofractal) to strong multifractal when temperature approaches to critical regime.
This indicates that the generalized Hurst exponent is a good indicator of phase transition along with 
statistical moments. The singularity spectrum and the complexity parameters have been employed to inspect multifractality level of 
magnetization time series at different temperatures. The shape of singularity spectrum around
critical point becomes very complicated which depict the complex dynamics of Ising system. The evolution of the complexity parameters
show the strong multifractality of the system at critical regime quantitatively.
The shuffling procedure has identified the sources of multifractality of the system near critical point stem both from
broad probability density function and long-term correlations.
The visibility graph method has been employed to convert the magnetization time series to complex networks.
Heterogeneous degree distributions of the complex networks at three temperature regimes have
shown the fractal nature of magnetization time series. Basic topological quantities of networks 
can capture the geometrical variation of time series. 
The decreases and increases of those topological quantities near critical regime have manifested the
critical dynamics of Ising system. Those evolution patterns can help us identifying how far the system is away form 
critical point. Thus we then conclude that the generalized Hurst exponent, the level of multifractality
and the topological quantities of visibility graphs can serve as early warnings for diverse of complex systems. 
The MF-DFA and the visibility graph method may not only be limited here for the analysis of two-dimensional Ising system, but can be
used as powerful tools to analyze the critical behaviours of many other systems. Thus we propose that the
multifractality and complex network topological quantities can be used as  early warnings 
for various complex systems near critical thresholds.
\section*{Acknowledgments}
This work is supported in part by the Programme of Introducing Talents of Discipline
to Universities under grant NO. B08033.


%
%
%

\bibliographystyle{plain}
\bibliography{ising}

\end{document}